\documentclass[letterpaper]{article}

\usepackage[T1]{fontenc}

\usepackage{geometry}
\usepackage{setspace}
\usepackage{romannum}
\usepackage{achemso}

\usepackage{graphicx}
\usepackage{float}
\newfloat{scheme}{htbp}{los}
\floatname{scheme}{Scheme}
\floatname{chart}{Chart}
\newfloat{graph}{htbp}{loh}

\usepackage{chemformula} 
\usepackage[version = 4]{mhchem} 

\usepackage{authblk}
\author[1]{Igor V.  Parshin}
\author[1]{Igor V.  Rubtsov}
\author[1]{Alexander L.  Burin*}
\affil[1]{Department of Chemistry, Tulane University}

\title{Thermal conductivity of aligned polymers with kinks}
\date{*Email: aburin@tulane.edu}

\begin{document}

\maketitle

\begin{abstract}
Thermal conductivity of aligned polymer molecules can be exceptionally high along the alignment direction due to energy transport through strong covalent bonds.  At the same time, it is highly sensitive to molecular conformation, varying by orders of magnitude as a result of gauche kinks.
Here, we theoretically investigate phonon transport in kinked polymers by numerically evaluating thermal conductivity and interpreting the results in terms of phonon scattering from randomly distributed kinks. For strongly aligned polymers with restricted deviations from a linear backbone, we find that heat transport becomes superdiffusive at long lengths, with thermal conductivity scaling as $\kappa \propto L^{1/3}$. 
At shorter lengths, thermal conductivity exhibits non-monotonic behavior: it increases at very short scales due to ballistic transport of almost all phonons, then decreases at intermediate lengths due to the Anderson localization of  most phonon modes.  These results are consistent with experiments and molecular dynamics simulations, and they elucidate the microscopic mechanisms governing heat transport in polymers.  
\end{abstract}

\noindent
{\bf Keywords:} Thermal conductivity, Polymer,  gauche kinks, Anderson localization,  Superdiffusive transport




\noindent
{\bf Introduction.} Thermal conductivity of aligned polymer molecules can be very high along the alignment direction, in some cases rivaling or exceeding that of metals due to strong covalent bonding  \cite{1977ChoyReviewPolymerThCond,1988PolymHighThCond,2004Thermcondassembled,2010PolyethNanoFibHigThCond,2010MDCondSensFluctShort,2012MolChainMDThC,2013FibersHighThCond,2017HighThmCLattprl,2019PolymerHighThCond,2020MDLangEnergyTrNitzan,2021VdVAnisotropic,2021MDPolyethAnisLiu,2022QuantPhTrDvira,2024YoungModulusAndThermCond}.  This expectation is consistent with diamond, one of the best-known thermal conductors, which is composed of similar covalent bonds \cite{2022highthcond-diam-nanow,2004DiamondSpeedofSound}.   While atoms in diamond are bonded nearly isotropically, alkyl chains exhibit strong anisotropy, leading to highly directional heat transport. Materials with such high thermal conductivities are of significant interest for applications in thermal management, heat exchangers, and energy conversion  \cite{1996RevUlmanSAM,1998QuantThermCond,1998HeatTranspRev,2000NatQuantThermCond,2004TechMolRev,DlottScience07,AbeScience07,SegalReview2016,2017RevEnTrInMolJ,XingThTrReview18,
ab19IgorReview,2021PhononEngReview,2021ballist,
2021ReviewFromnanowtosupercond,2022QuantPhTrDvira,2023RevNonFourPhTranspLiviLepri,2023MDSimulRev,2023HeatTrModNeuralNetw,2024EnergConvTranspRev,2024RevThCondControlLiu}.  Moreover, understanding vibrational energy transport in molecules and biomolecules is essential for controlling reaction kinetics, as energy transfer is mediated by molecular vibrations  \cite{2004-VibrEnReactGruebele,HammBotan07peptides,PandeyLeitner2016ThermSign,LeitnerReviewProtein18,
LeitnerBook,Leitner2019EnTransBiomol,
Elenewski2019BiomolTr,2019VibrFlReact,2019VibrReactControl,Stock2022EnTranspOptProt,2024ThCondProtLeitner,2022BarrCrossProt,2024LeitnerReact,2025VibrDynPept}.


Thermal transport in long one-dimensional conductors is governed by the competition between two mechanisms. On the one hand, long-wavelength phonons propagate ballistically and always exist as Goldstone modes arising from translational symmetry  \cite{1962GoldstoneSymmPart,15LeitnerReview,Lebowitz15,2021RevLiviPedagog,2021ReviewFromnanowtosupercond}.  On the other hand, scattering by defects suppresses transport and leads to Anderson localization of phonons  \cite{Anderson58,Abrahams1979ScThLoc}.  The interplay of these effects results in superdiffusive heat transport, characterized by thermal conductivity that increases with system length.  At short lengths, conductivity increases linearly, corresponding to the ballistic (quantum) regime
   \cite{1998QuantThermCond,1998HeatTranspRev}.  At longer lengths, it follows a power law $\kappa \propto L^{\beta}$, $0 < \beta <1$,  where $\beta=1/2$ for longitudinal phonon scattering by defects \cite{Lebowitz67,20011DGlass}, $\beta=1/3$ or $2/5$ for anharmonic scattering \cite{1998LepriAnomalThCondFPUCl,Turbulence02}, and $\beta=1/3$ for longitudinal-to-transverse phonon scattering by defects \cite{ab25SuperDiffPh}.  These predictions are broadly consistent with experimental observations in one-dimensional systems \cite{2000NatQuantThermCond,2017cntHighKappa_mmlong,2017BallTr4KSiNanow033} including polymer chains such as polyethylene \cite{MeierMolChains14} ($4$–$18$ carbon atoms, $\beta \sim 0.38 \pm 0.2$) and poly(methyl acrylate) (PMA) \cite{2024prlnanopartonpolymsuperdiff} ($300$–$1000$ monomers, $\beta \approx 0.46$).  

Despite this general agreement, the microscopic origin of the observed thermal transport in molecular systems remains unclear \cite{MeierMolChains14,2024prlnanopartonpolymsuperdiff}. For example, the weak temperature dependence of thermal conductivity in polyethylene \cite{MeierMolChains14} suggests that defect scattering dominates over anharmonic effects, since the latter would lead to decreasing conductivity with increasing temperature.   In chemically pure polymers, the most relevant structural defects are gauche kinks \cite{1969florystatistical,2003rubinsteinpolymer}. Molecular dynamics simulations \cite{2019ThermCondKinksXuhui,2020NitzanHeatCondMDRegimes} indicate that kinks can strongly suppress superdiffusive transport, restoring Fourier behavior (length-independent conductivity), which is not observed experimentally. Furthermore, in PMA, thermal conductivity has been reported to decrease at very long chain lengths, contradicting existing theoretical expectations. These discrepancies highlight the need for a deeper theoretical understanding of heat transport in kinked polymer systems.

\begin{figure}
  \centering
   \includegraphics[width=\columnwidth]{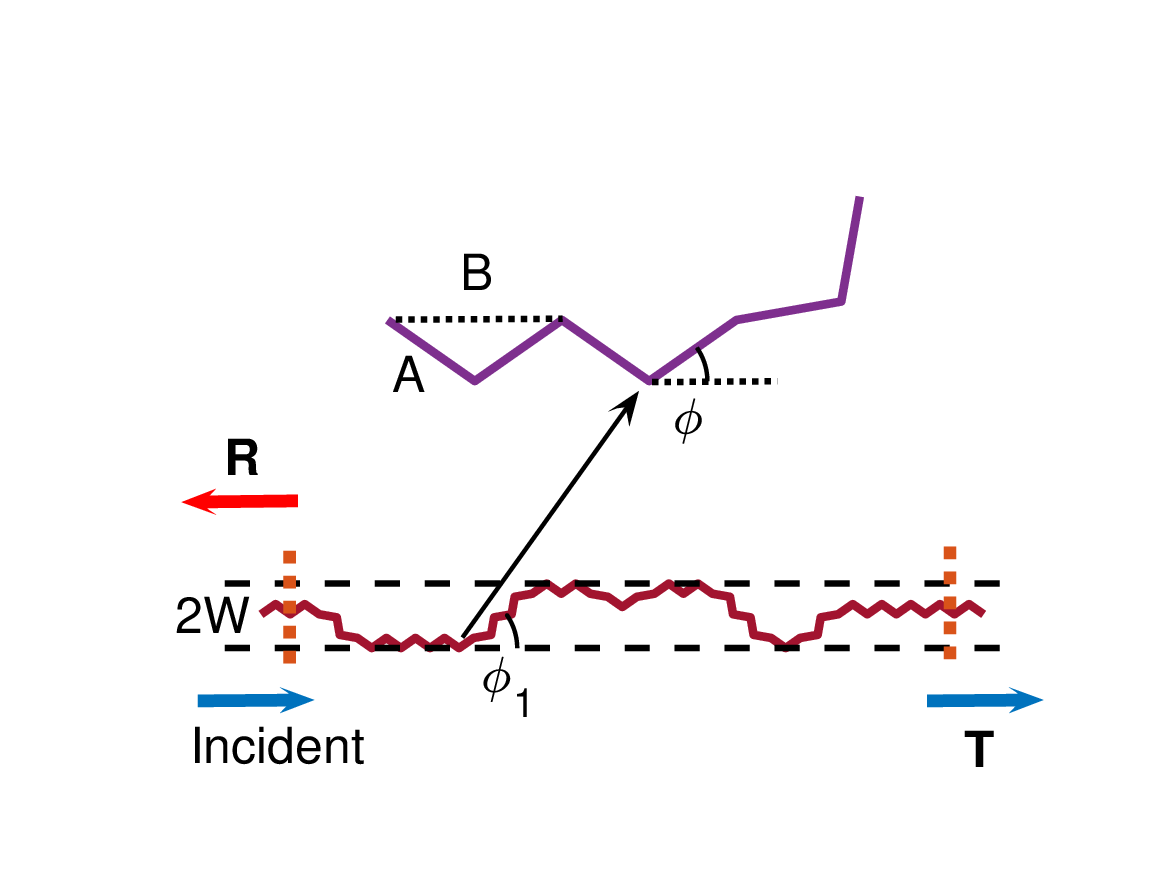}
\caption{The fence model of polymer molecule with kinks.  Displacement of $y$-coordinated is restricted by the width $2W$.  Thick dashed lines indicate molecular borders. Thick arrows show incident, transmitted and reflected waves characterized by their transmission and reflection coefficients $\mathbf{T}$ and $\mathbf{R}$.  The top chart shows the enhanced fragment of the molecule containing kink. }
  \label{fig:kinkpic}
\end{figure}

\begin{figure}
  \centering
   \includegraphics[width=\columnwidth]{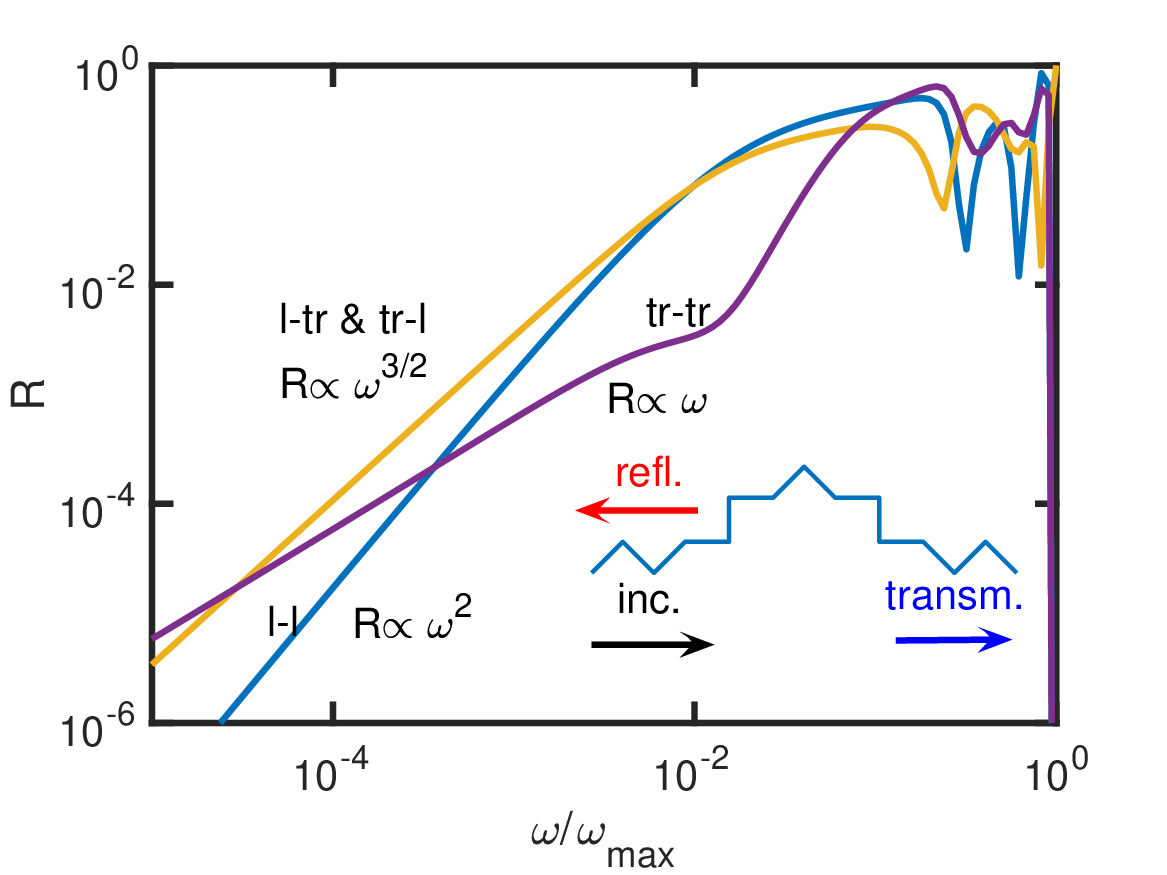}
\caption{Reflection from a set of kinks modeling elementary defect.}
  \label{fig:elem-refl}
\end{figure}

In this work, we investigate length-dependent phonon thermal conductivity in aligned polymer chains containing randomly distributed gauche kinks with limited transverse displacements (Fig.~\ref{fig:kinkpic}). We show that thermal conductivity exhibits a non-monotonic dependence on chain length. At short lengths, it increases due to ballistic transport of thevast majority of phonons. At intermediate lengths, it decreases as Anderson localization suppresses transport for the majority of phonons. At long lengths, thermal conductivity increases again, following the universal superdiffusive scaling $\kappa \propto L^{1/3}$, governed by longitudinal-to-transverse phonon scattering \cite{ab25SuperDiffPh} and dominated by the ballistic propagation of long-wavelength modes (Fig.~\ref{fig:elem-refl}).

{\bf Model and Methods.} We model the polymer chain using the two-dimensional fence model introduced in Ref.~\cite{ab25SuperDiffPh}, extended here to include kinks that capture the structure of alkane chains (Fig.~\ref{fig:kinkpic}). The fence geometry is defined by the angle $\phi$ between the bonds and the molecular axis in the absence of kinks, while kinks are introduced as local rotations of the molecular axis by  angle $\phi_1$.

Harmonic interactions between neighboring and next-nearest neighboring atoms are characterized by force constants $A$ and $B$, respectively. These interactions are introduced in a translationally and rotationally invariant form as $V_{ij}=C((\mathbf{u}_i-\mathbf{u}_j)\mathbf{n}_{ij})^2/2$, where $C$ denotes the force constant ($A$ or $B$),  $\mathbf{u}_i$ is the displacement of the atom $i$  and $\mathbf{n}_{ij}$ is the unit vector connecting sites $i$ and $j$.  Based on density functional theory (DFT) calculations of the Hessian matrix for alkane chains with kinks, we use $A \approx 4.8B$ and $B \approx 0.0244$ a.u  \cite{Gaussian}. 

\begin{figure}
  \centering
   \includegraphics[width=\columnwidth]{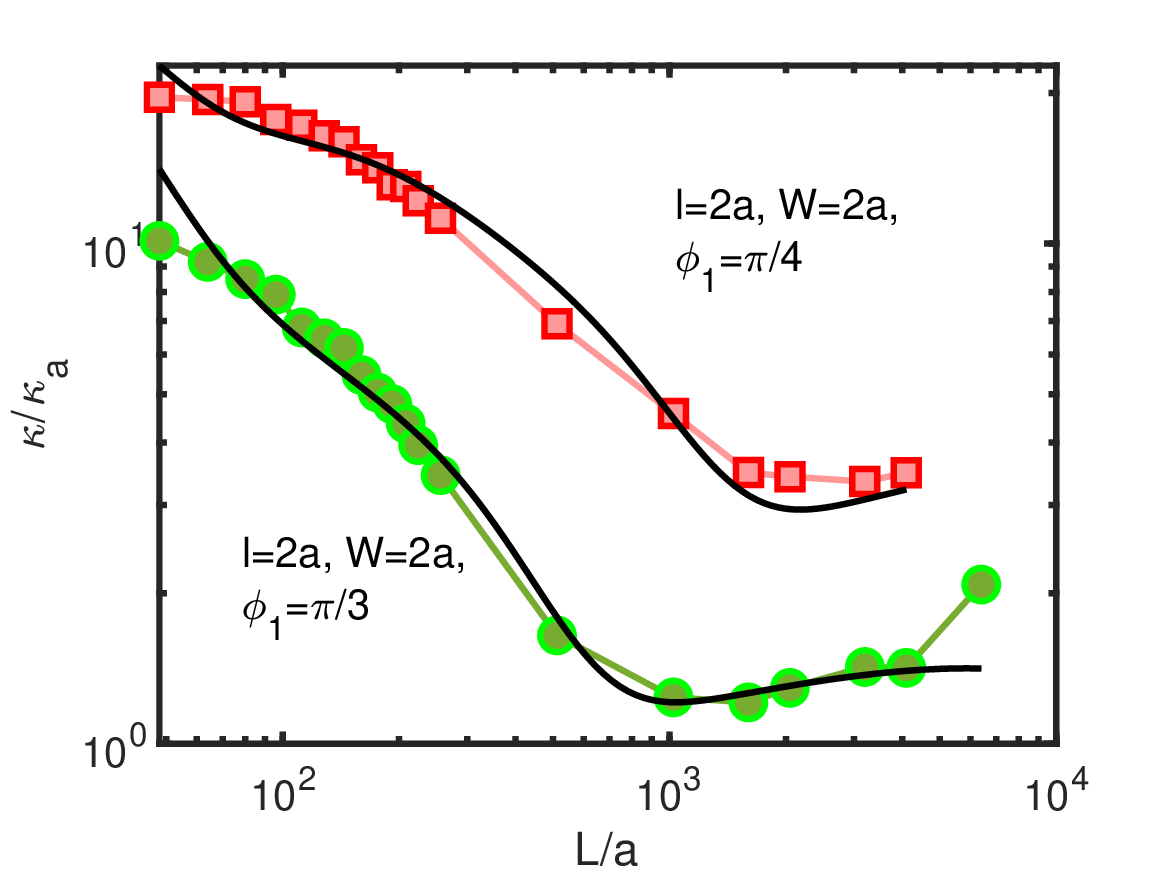} 
\caption{Interpolation of a thermal conductivity constructed using the elementary kink model.  
} 
  \label{fig:KappaKinksInt}
\end{figure}

We consider $\phi = 0.6109$ rad ($35^\circ$), corresponding to an inter-bond angle of $2(90^\circ - \phi) = 110^\circ$ simulating the C-C-C angle like in alkane chains.   For the kink angle, we use $\phi_1 = \pi/4$, which is close to the physically relevant value, and also $\phi_1 = \pi/3$ to access the long-length asymptotic regime of thermal conductivity (Fig.~\ref{fig:KappaKinksInt}), which requires significantly longer chains for $\phi_1 = \pi/4$ than for $\phi_1=\pi/3$.

Kinks are introduced as local turns of the molecular axis by angles $\pm \phi_1$ at randomly selected sites. The distances between successive kinks are generated from a Poisson distribution with a fixed mean inter-kink distance $l$, while the total number of kinks $k$ is kept constant. To constrain transverse fluctuations, configurations in which the atom vertical  displacement exceeds the allowed interval $(-W, W)$ are discarded. To model aligned systems, we impose a unidirectional constraint by limiting the cumulative deviation of the molecular axis within $\pm \phi_1$. A typical configuration for $W = a$ (where $a$ is the period of the kink-free chain) is shown in Fig.~\ref{fig:kinkpic}. We consider two representative regimes: ($l = a$,  $W = a$) and ($l = 2a$,  $W = 2a$). The resulting thermal conductivity is highly sensitive to these parameters; a systematic analysis of this dependence will be presented elsewhere.

Since thermal conductivity is determined by phonon transmission through the molecule, we connect each generated atomic chain to two kink-free semi-infinite leads on the left and right, matched at identical transverse coordinates. This simplified setup neglects the detailed structure of the molecule-lead junctions. However, for sufficiently long molecules, the intrinsic chain resistance is expected to dominate the total thermal resistance, making the junction details less critical. This assumption is supported by experimental observations showing that thermal conductivity increases with molecular length \cite{MeierMolChains14,2024prlnanopartonpolymsuperdiff}, indicating that transport is primarily governed by the molecule itself.

We neglect anharmonic interactions, assuming that the temperature is sufficiently low and the molecular length sufficiently short. This approximation is consistent with experimental data at temperatures up to room temperature (e.g., Refs.~\cite{MeierMolChains14,2017BallTr4KSiNanow033}), where thermal conductivity is weakly temperature dependent or even increases with temperature, suggesting that scattering on defects dominates over anharmonic effects.

The kink-free fence model supports two acoustic phonon branches: longitudinal (LA) and transverse (TA). In the long-wavelength limit (small wavevector $q$), their dispersions are linear, $\omega_l = cq$, and quadratic, $\omega_t = \tan(\phi)c a q^2$, respectively \cite{landau1986Elasticitytheory,ab25SuperDiffPh}. Here $c = a\sqrt{B/M} \approx 1.6 \times 10^4$ m$/$s is the sound velocity and $M = 14$ a.m.u for a methylene group. This estimate is consistent with previous results for alkane chains based on both DFT and Morse potentials \cite{ab2023Cherenkov,SegalNitzan03}.

For simplicity, we assume that the temperature is sufficiently high such that the thermal energy satisfies $k_B T \gg \hbar c/a$, which is qualitatively valid near room temperature. In this regime, the thermal conductivity can be expressed using the generalized Landauer formula \cite{Landauer57Classic,1988EquatForThermCond,1998QuantThermCond,1998HeatTranspRev,SegalNitzan03}, assuming a small temperature difference between the leads 
\begin{eqnarray}
\kappa = \kappa_{a} \frac{L}{a\omega_{\rm max}}\int_0^{\omega_{\rm max}}d\omega  \mathcal{T}(\omega),  ~ \kappa_{a}=\frac{k_B a\omega_{\rm max}}{2\pi}, 
\nonumber\\
\mathcal{T} = T_{\rm ll}+T_{\rm lt}+T_{\rm tl}+T_{\rm tt},  
\label{eq:kappaDvira}
\end{eqnarray}  
where the transmission tensor $T_{\mu\nu}$ determines the transmission of incident wave $\mu$ (LA or TA) forming the transmitted wave $\nu$ and $\omega_{\rm max}\approx 1.4809c/a$ stands for acoustic phonon bandwidth identical for longitudinal and transverse phonons. 

Phonon transmission coefficients are evaluated using the advanced transfer matrix method \cite{ab25SuperDiffPh,2016GeneralizedTransfMatr1D}. We compute the relevant blocks of the Hessian matrix: the coupling between the molecule and the left lead, $\widehat{V}_l$, the molecular Hessian projected onto internal coordinates, $\widehat{H}_{\rm in}$, and the coupling to the right lead, $\widehat{V}_r$. The effective lead–lead coupling is then expressed through the matrix products $V_{l,r}^{\dagger}(-\omega^2\widehat{I}+\widehat{H}_{\rm in})^{-1}V_{l,r}$ where $\widehat{I}$ is the identity matrix. Using these quantities together with representations of incident, reflected, and transmitted phonon modes, we compute all transmission and reflection coefficients.  

We verified that this approach yields results identical to those obtained using standard Green’s function methods \cite{1971TrThrGFCaroli,1994MWCoductMR,2004TransmGFRatner,2008QuantThermTranspRev,XingThTrReview18}. However, the transfer matrix method is computationally more efficient and exhibits superior convergence at low frequencies, allowing reliable calculations down to $\omega \sim 10^{-6}\omega_{\rm max}$. This was confirmed by verifying energy conservation, $\sum_{\mu}(T_{\mu\nu} + R_{\mu\nu}) = 1$.

At frequencies lower than $10^{-6}\omega_{\rm max}$, numerical precision limits become significant, as $\omega^2$ approaches the machine precision threshold $\sim 10^{-16}\omega_{\rm max}^2$. This limitation restricts the maximum number of CH$_2$ groups   to approximately $10^4$.  Accordingly, we use a lower integration cutoff $\omega_{\rm min} = 10^{-6}\omega_{\rm max}$ in Eq.~(\ref{eq:kappaDvira}), which does not substantially affect the results.

The integrals are evaluated using one-dimensional numerical integration in MATLAB \cite{MATLAB:2019}. Results are averaged over several hundred random realizations for short chains and approximately ten realizations for the longest chains.

\begin{figure}
  \centering
   \includegraphics[width=\columnwidth]{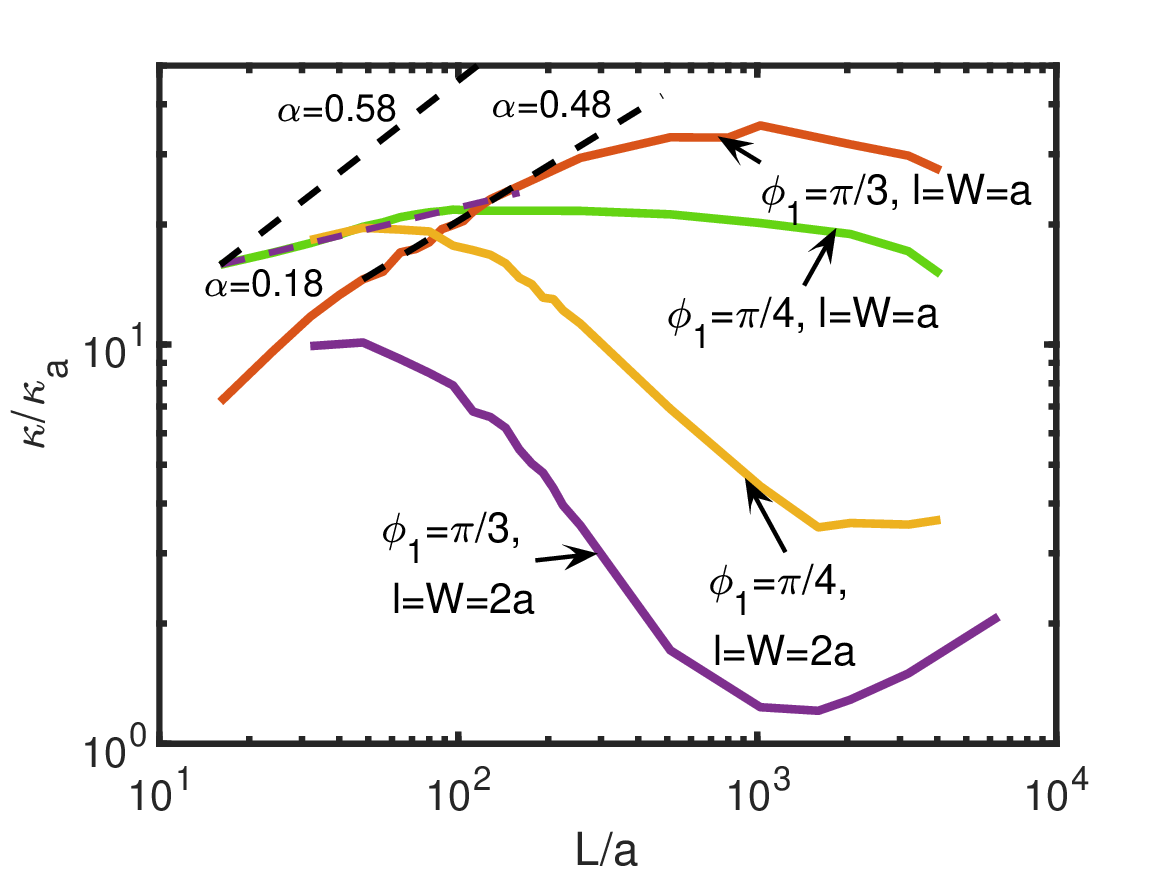}
\caption{Average thermal conductivity vs length for different realizations  of kinks with restricted transverse displacement, Fig. \ref{fig:kinkpic}.  Straight black dashed line shows $L^{0.38\pm 0.2}$ and  $L^{0.46}$ length  dependencies  discovered in Refs.  \cite{MeierMolChains14,2024prlnanopartonpolymsuperdiff}  for comparison. }
  \label{fig:KappaKinks}
\end{figure}

\noindent
{\bf Results.} The length dependence of the average thermal conductivity is shown in Fig.~\ref{fig:KappaKinks} for atomic chains generated with two different kink angles $\phi_1$ and two sets of parameters for the inter-kink separation $l$ and transverse displacement $W$. We found that thermal conductivity exhibits a non-monotonic dependence on chain length, which can be divided into three distinct regimes. It increases at short lengths, decreases by approximately a factor of four at intermediate lengths, and increases again at the longest lengths.

The initial increase is less pronounced for chains with the largest transverse width ($W = 2a$), which exhibit the lowest overall thermal conductivity. Conversely, the long-length increase is not clearly observed for the most conductive configurations with $W = a$.

The characteristic scale of molecular thermal conductivity in our model is $\kappa_a \approx 5.1 \times 10^{-20}$ m·pW/K (see Eq.~(\ref{eq:kappaDvira})), while the experimental value reported in Ref.~\cite{MeierMolChains14} is approximately $3 \times 10^{-20}$ m·pW/K. The discrepancy becomes more significant at short lengths relevant to the experiment. This difference is likely due to a limitation of the present two-dimensional fence model, which includes only a single transverse phonon branch. Incorporating an additional transverse mode is expected to reduce the thermal conductivity and improve agreement with experiment. It can also be connected to the effect of junctions contributing through the additional small factor  expressing transmission from the lead to the  molecule which is unity in the present model. 

At short lengths and small transverse width ($W = a$), the length dependence is consistent with experimentally observed scaling laws, $L^{0.38 \pm 0.2}$ and $L^{0.46}$, reported in Refs.~\cite{MeierMolChains14,2024prlnanopartonpolymsuperdiff}. At the longest lengths, thermal conductivity decreases with increasing length, in agreement with the behavior observed in Ref.~\cite{2024prlnanopartonpolymsuperdiff} at comparable scales.

\noindent
{\bf Discussion.}  To interpret the observed behavior, we model phonon scattering as arising from a set of independent, non-intersecting kink-like defects with density $n$. Each defect consists of four kinks with turn angles $\phi_{\rm 1,mod}$, $-\phi_{\rm 1,mod}$, $-\phi_{\rm 1,mod}$, and $\phi_{\rm 1,mod}$ (see Fig.~\ref{fig:elem-refl}), where $\phi_{\rm 1,mod}$ serves as an adjustable parameter controlling the overall defect strength. In addition, we introduce a second parameter, $\phi_{\rm mod}$, which defines the angle between the chain axis and the bond direction within the defect, thereby tuning the local geometric distortion and effective stiffness. Outside the defect, the chain retains its original geometry. The length of the chain segments between adjacent kinks within a defect is chosen to be two lattice periods, ensuring that the chain before and after the defect remains at the same vertical position. Use of one lattice period does not fit well the calculation results.  This construction models defects that preserve the overall transverse constraint of the chain.

The scattering of a phonon with frequency $\omega$ by an individual defect is characterized by a reflection coefficient $R(\omega)$ (see Fig.~\ref{fig:elem-refl}). At long wavelengths, translational invariance requires that reflection vanishes as a power law, $R(\omega) \propto \omega^{\gamma}$ with power low exponent $\gamma$ depending on the specific scattering mechanism as illustrated in Fig. \ref{fig:elem-refl}.  

In one dimension, where Anderson localization dominates~\cite{Abrahams1979ScThLoc}, the transmission through a chain of length $L$ with randomly distributed, identical defects can be approximated in the dilute, uncorrelated limit by \cite{1984DOROKHOV381}
\begin{eqnarray}
T(\omega) \approx \frac{1}{\cosh\left(n L R(\omega)/2\right)^2}.
\label{eq:TransmAn}
\end{eqnarray}
This expression provides a unified framework for understanding the different transport regimes observed in our calculations.

For short chains, such that $n L R(\omega) \leq 1$ over the frequency range contributing most to heat transport, transmission remains close to unity for most of the phonons  and transport is predominantly ballistic, leading to an increase of thermal conductivity with length. At intermediate lengths, the condition $n L R(\omega) > 1$ is satisfied for the majority of relevant phonon modes, resulting in exponential suppression of transmission and a corresponding decrease in thermal conductivity due to Anderson localization.

In the limit of very long chains, transport is dominated by a narrow low-frequency window where $n L R(\omega) < 1$, i.e., by long-wavelength phonons that are only weakly scattered. In this regime, the thermal conductivity exhibits superdiffusive scaling, 
 $\kappa \propto L^{\beta}$, where $\beta = \frac{1 - \gamma}{\gamma}$,
which follows directly from the universal low-frequency behavior $R(\omega) \propto \omega^{\gamma}$  \cite{20011DGlass}.  

We model a representative defect as a four-kink bridge, shown in Fig.~\ref{fig:elem-refl} (lower right). This construction provides a minimal model of disorder induced by kinks under the constraints of limited transverse displacement and partial alignment of the molecular axis. The defect geometry is parameterized by the angles $\phi_d$ and $\phi_{1d}$ (see Fig.~\ref{fig:kinkpic}), while the defect density $n$ is treated as an additional fitting parameter. These parameters are determined by minimizing the deviation between thermal conductivities obtained from exact transmission calculations (Fig.~\ref{fig:KappaKinks}) and those computed using the approximate transmission formula in Eq.~(\ref{eq:TransmAn}). Reflection coefficients entering Eq.~(\ref{eq:TransmAn}) are evaluated using the full frequency-dependent scattering amplitudes for both longitudinal and transverse phonons.

The frequency dependence of the reflection coefficient for a representative defect with $\phi_d = \phi_{1d} = \pi/4$ is shown in Fig.~\ref{fig:elem-refl}. It is consistent with earlier results for phonon scattering by force-constant defects \cite{ab25SuperDiffPh}, exhibiting the expected suppression at low frequencies.

We perform a Monte Carlo optimization of the defect parameters to reproduce the numerically obtained thermal conductivity for chains with $W = 2a$ (chains with $W = a$ are excluded, as the long-length regime is not reached). This procedure yields a good agreement for parameter sets $\phi_{\rm mod} = 0.7615$, $\phi_{\rm 1,mod} = 1.026$, and $n = 0.0325$ for $\phi_1 = \pi/4$, and $\phi_{\rm mod} = 0.8042$, $\phi_{\rm 1,mod} = 1.0153$, and $n = 0.057$ for $\phi_1 = \pi/3$, as shown in Fig.~\ref{fig:KappaKinksInt}.

The remaining discrepancies at short lengths likely arise from the specifics of the chain generation procedure, which fixes the number of kinks, whereas Eq.~(\ref{eq:TransmAn}) assumes an ensemble of all possible configurations, including defect-free chains. Such configurations can contribute significantly at short lengths.

At very long lengths, thermal transport is expected to be dominated by longitudinal phonons, leading to the universal scaling $\kappa \propto L^{1/3}$ \cite{ab25SuperDiffPh}. This regime is not fully reached in our simulations due to system size limitations. At intermediate lengths, however, transverse phonons provide the dominant contribution to heat transport.

The experimentally observed scaling $\kappa \propto L^{0.38 \pm 0.2}$ in Ref.~\cite{MeierMolChains14} likely does not originate from the long-wavelength phonons responsible for asymptotic transport at much larger scales. Instead, it reflects a competition between Anderson localization, which suppresses thermal conductivity, and the increasing number of contributing phonon modes with system length, primarily associated with longitudinal phonons. A similar mechanism may underlie the behavior observed in poly(methyl acrylate) (PMA) \cite{2024prlnanopartonpolymsuperdiff}, which qualitatively resembles our results for the case $l = W = a$ (Fig.~\ref{fig:KappaKinks}), including the increase $\kappa \propto L^{0.46}$ for $L < 1000a$ followed by a decrease at larger lengths. However, this comparison remains qualitative, as the present fence model does not capture the full structural complexity of PMA.

 According to Ref.  ~\cite{MeierMolChains14} the observed behavior of thermal conductivity in alkane chains can be explained based on the results of Ref.  \cite{SegalNitzan03}, where the molecule was considered defect-free and the thermal resistance is originated from the molecule-lead junctions.  It is not clear for us whether defect-free model is justified since at room temperature there is around one gauche kink per three carbon atoms in equilibrium, that might be, however,  broken by the specific experimental conditions.   The present model suggests that the increase in molecular length should lead to reduction in thermal conductivity.   Thus we propose to investigate thermal conductivity of alkane chains at longer lengths to validate the proposed explanation.

\noindent
{\bf Conclusions.} We investigated heat transport through acoustic phonon bands in aligned polymer chains containing kinks using a simplified fence model that captures essential features of alkane-like systems. We find that thermal conductivity exhibits a complex non-monotonic dependence on chain length, arising from the interplay between ballistic transport and Anderson localization.

At short and intermediate lengths, all phonon modes contribute to transport, leading first to an increase and then to a decrease in thermal conductivity with increasing chain length. This behavior is qualitatively consistent with experimental observations reported in Refs.~\cite{MeierMolChains14,2024prlnanopartonpolymsuperdiff}.   If this interpretation is correct for polyethylene, it would further imply a decrease in thermal conductivity with increasing molecular length at sufficiently long scales.   Such decrease is observed in PMA  \cite{2024prlnanopartonpolymsuperdiff},  where our consideration is applicable only qualitatively.  

At the longest lengths, heat transport is expected to be governed by long-wavelength phonons, eventually giving rise to the asymptotic scaling $\kappa \propto L^{1/3}$ associated with longitudinal phonon transport. However, in the intermediate regime of hundreds to thousands of bond lengths, transport is dominated by transverse phonons, leading to a weaker effective length dependence. In this regime, thermal transport is significantly slower than the sound propagation time,  that is qualitatively consistent with the observations in Ref.~\cite{DlottScience07}.  Nevertheless, since this behavior appears  at relatively short lengths, the dominant role of transverse phonons in real systems remains an open question.

We also find that phonon transport is highly sensitive to the transverse displacement $W$ induced by kinks and to the molecular axis rotation angle $\phi_1$ at kink sites.  Increasing either parameter leads to a strong suppression of thermal conductivity, suggesting a practical mechanism for controlling heat transport via molecular alignment.  This observation is qualitatively consistent with the thermal conductivity dramatic increase in polyethylene nano-fibers due to alignment \cite{2010PolyethNanoFibHigThCond}.  
The detailed dependence of thermal conductivity on these parameters is left for future studies.  Yet, it is noticeable that in addition to the number of kinks, the transverse displacements of atoms $W$  represents another important parameter controlling molecular thermal conductivity.

It is interesting that according to the present consideration $L\beta - L\alpha$ phase transition in membranes accompanied by lipid misalignment and numerous kink formations \cite{ab2023Membr} should lead to the substantial reduction of membrane thermal conductivity that is interesting to probe using the methods of two-dimensional infra-red spectroscopy \cite{ab19IgorReview}.

\noindent
{\bf Acknowledgements.}  This work is supported by the National Science Foundation
(Grant No. CHE-2201027).

\bibliography{Vibr1.bib}

\providecommand{\latin}[1]{#1}
\makeatletter
\providecommand{\doi}
  {\begingroup\let\do\@makeother\dospecials
  \catcode`\{=1 \catcode`\}=2 \doi@aux}
\providecommand{\doi@aux}[1]{\endgroup\texttt{#1}}
\makeatother
\providecommand*\mcitethebibliography{\thebibliography}
\csname @ifundefined\endcsname{endmcitethebibliography}
  {\let\endmcitethebibliography\endthebibliography}{}
\begin{mcitethebibliography}{83}
\providecommand*\natexlab[1]{#1}
\providecommand*\mciteSetBstSublistMode[1]{}
\providecommand*\mciteSetBstMaxWidthForm[2]{}
\providecommand*\mciteBstWouldAddEndPuncttrue
  {\def\EndOfBibitem{\unskip.}}
\providecommand*\mciteBstWouldAddEndPunctfalse
  {\let\EndOfBibitem\relax}
\providecommand*\mciteSetBstMidEndSepPunct[3]{}
\providecommand*\mciteSetBstSublistLabelBeginEnd[3]{}
\providecommand*\EndOfBibitem{}
\mciteSetBstSublistMode{f}
\mciteSetBstMaxWidthForm{subitem}{(\alph{mcitesubitemcount})}
\mciteSetBstSublistLabelBeginEnd
  {\mcitemaxwidthsubitemform\space}
  {\relax}
  {\relax}

\bibitem[Choy(1977)]{1977ChoyReviewPolymerThCond}
Choy,~C. \emph{Polymer} \textbf{1977}, \emph{18}, 984--1004\relax
\mciteBstWouldAddEndPuncttrue
\mciteSetBstMidEndSepPunct{\mcitedefaultmidpunct}
{\mcitedefaultendpunct}{\mcitedefaultseppunct}\relax
\EndOfBibitem
\bibitem[Kanamoto \latin{et~al.}(1988)Kanamoto, Tsuruta, Tanaka, Takeda, and
  Porter]{1988PolymHighThCond}
Kanamoto,~T.; Tsuruta,~A.; Tanaka,~K.; Takeda,~M.; Porter,~R.~S.
  \emph{Macromolecules} \textbf{1988}, \emph{21}, 470--477\relax
\mciteBstWouldAddEndPuncttrue
\mciteSetBstMidEndSepPunct{\mcitedefaultmidpunct}
{\mcitedefaultendpunct}{\mcitedefaultseppunct}\relax
\EndOfBibitem
\bibitem[Ge \latin{et~al.}(2004)Ge, Cahill, and Braun]{2004Thermcondassembled}
Ge,~Z.; Cahill,~D.~G.; Braun,~P.~V. \emph{The Journal of Physical Chemistry B}
  \textbf{2004}, \emph{108}, 18870--18875\relax
\mciteBstWouldAddEndPuncttrue
\mciteSetBstMidEndSepPunct{\mcitedefaultmidpunct}
{\mcitedefaultendpunct}{\mcitedefaultseppunct}\relax
\EndOfBibitem
\bibitem[Shen \latin{et~al.}(2010)Shen, Henry, Tong, Zheng, and
  Chen]{2010PolyethNanoFibHigThCond}
Shen,~S.; Henry,~A.; Tong,~J.; Zheng,~R.; Chen,~G. \emph{Nature Nanotechnology}
  \textbf{2010}, \emph{5}, 251--255\relax
\mciteBstWouldAddEndPuncttrue
\mciteSetBstMidEndSepPunct{\mcitedefaultmidpunct}
{\mcitedefaultendpunct}{\mcitedefaultseppunct}\relax
\EndOfBibitem
\bibitem[Hu \latin{et~al.}(2010)Hu, Zhang, Hu, Wang, Li, and
  Keblinski]{2010MDCondSensFluctShort}
Hu,~L.; Zhang,~L.; Hu,~M.; Wang,~J.-S.; Li,~B.; Keblinski,~P. \emph{Phys. Rev.
  B} \textbf{2010}, \emph{81}, 235427\relax
\mciteBstWouldAddEndPuncttrue
\mciteSetBstMidEndSepPunct{\mcitedefaultmidpunct}
{\mcitedefaultendpunct}{\mcitedefaultseppunct}\relax
\EndOfBibitem
\bibitem[Liu and Yang(2012)Liu, and Yang]{2012MolChainMDThC}
Liu,~J.; Yang,~R. \emph{Phys. Rev. B} \textbf{2012}, \emph{86}, 104307\relax
\mciteBstWouldAddEndPuncttrue
\mciteSetBstMidEndSepPunct{\mcitedefaultmidpunct}
{\mcitedefaultendpunct}{\mcitedefaultseppunct}\relax
\EndOfBibitem
\bibitem[Wang \latin{et~al.}(2013)Wang, Ho, Segalman, and
  Cahill]{2013FibersHighThCond}
Wang,~X.; Ho,~V.; Segalman,~R.~A.; Cahill,~D.~G. \emph{Macromolecules}
  \textbf{2013}, \emph{46}, 4937--4943\relax
\mciteBstWouldAddEndPuncttrue
\mciteSetBstMidEndSepPunct{\mcitedefaultmidpunct}
{\mcitedefaultendpunct}{\mcitedefaultseppunct}\relax
\EndOfBibitem
\bibitem[Shulumba \latin{et~al.}(2017)Shulumba, Hellman, and
  Minnich]{2017HighThmCLattprl}
Shulumba,~N.; Hellman,~O.; Minnich,~A.~J. \emph{Phys. Rev. Lett.}
  \textbf{2017}, \emph{119}, 185901\relax
\mciteBstWouldAddEndPuncttrue
\mciteSetBstMidEndSepPunct{\mcitedefaultmidpunct}
{\mcitedefaultendpunct}{\mcitedefaultseppunct}\relax
\EndOfBibitem
\bibitem[Xu \latin{et~al.}(2019)Xu, Kraemer, Song, Jiang, Zhou, Loomis, Wang,
  Li, Ghasemi, Huang, Li, and Chen]{2019PolymerHighThCond}
Xu,~Y.; Kraemer,~D.; Song,~B.; Jiang,~Z.; Zhou,~J.; Loomis,~J.; Wang,~J.;
  Li,~M.; Ghasemi,~H.; Huang,~X.; Li,~X.; Chen,~G. \emph{Nature Communications}
  \textbf{2019}, \emph{10}, 1771\relax
\mciteBstWouldAddEndPuncttrue
\mciteSetBstMidEndSepPunct{\mcitedefaultmidpunct}
{\mcitedefaultendpunct}{\mcitedefaultseppunct}\relax
\EndOfBibitem
\bibitem[Sharony \latin{et~al.}(2020)Sharony, Chen, and
  Nitzan]{2020MDLangEnergyTrNitzan}
Sharony,~I.; Chen,~R.; Nitzan,~A. \emph{The Journal of Chemical Physics}
  \textbf{2020}, \emph{153}, 144113\relax
\mciteBstWouldAddEndPuncttrue
\mciteSetBstMidEndSepPunct{\mcitedefaultmidpunct}
{\mcitedefaultendpunct}{\mcitedefaultseppunct}\relax
\EndOfBibitem
\bibitem[Kim \latin{et~al.}(2021)Kim, Mujid, Rai, Eriksson, Suh, Poddar, Ray,
  Park, Fransson, Zhong, Muller, Erhart, Cahill, and Park]{2021VdVAnisotropic}
Kim,~S.~E.; Mujid,~F.; Rai,~A.; Eriksson,~F.; Suh,~J.; Poddar,~P.; Ray,~A.;
  Park,~C.; Fransson,~E.; Zhong,~Y.; Muller,~D.~A.; Erhart,~P.; Cahill,~D.~G.;
  Park,~J. \emph{Nature} \textbf{2021}, \emph{597}, 660--665\relax
\mciteBstWouldAddEndPuncttrue
\mciteSetBstMidEndSepPunct{\mcitedefaultmidpunct}
{\mcitedefaultendpunct}{\mcitedefaultseppunct}\relax
\EndOfBibitem
\bibitem[He and Liu(2021)He, and Liu]{2021MDPolyethAnisLiu}
He,~J.; Liu,~J. \emph{Journal of Applied Physics} \textbf{2021}, \emph{130},
  225101\relax
\mciteBstWouldAddEndPuncttrue
\mciteSetBstMidEndSepPunct{\mcitedefaultmidpunct}
{\mcitedefaultendpunct}{\mcitedefaultseppunct}\relax
\EndOfBibitem
\bibitem[Gotsmann \latin{et~al.}(2022)Gotsmann, Gemma, and
  Segal]{2022QuantPhTrDvira}
Gotsmann,~B.; Gemma,~A.; Segal,~D. \emph{Applied Physics Letters}
  \textbf{2022}, \emph{120}, 160503\relax
\mciteBstWouldAddEndPuncttrue
\mciteSetBstMidEndSepPunct{\mcitedefaultmidpunct}
{\mcitedefaultendpunct}{\mcitedefaultseppunct}\relax
\EndOfBibitem
\bibitem[Zeng \latin{et~al.}(2024)Zeng, Liang, Zhang, Liu, Li, Lu, Han, Yao,
  Xu, Sun, and Li]{2024YoungModulusAndThermCond}
Zeng,~J.; Liang,~T.; Zhang,~J.; Liu,~D.; Li,~S.; Lu,~X.; Han,~M.; Yao,~Y.;
  Xu,~J.-B.; Sun,~R.; Li,~L. \emph{Small} \textbf{2024}, \emph{20},
  2309338\relax
\mciteBstWouldAddEndPuncttrue
\mciteSetBstMidEndSepPunct{\mcitedefaultmidpunct}
{\mcitedefaultendpunct}{\mcitedefaultseppunct}\relax
\EndOfBibitem
\bibitem[Liang \latin{et~al.}(2022)Liang, Xu, Han, Yao, Zhang, Zeng, Xu, and
  Wu]{2022highthcond-diam-nanow}
Liang,~T.; Xu,~K.; Han,~M.; Yao,~Y.; Zhang,~Z.; Zeng,~X.; Xu,~J.; Wu,~J.
  \emph{Materials Today Physics} \textbf{2022}, \emph{25}, 100705\relax
\mciteBstWouldAddEndPuncttrue
\mciteSetBstMidEndSepPunct{\mcitedefaultmidpunct}
{\mcitedefaultendpunct}{\mcitedefaultseppunct}\relax
\EndOfBibitem
\bibitem[Wang \latin{et~al.}(2004)Wang, Hsu, Pu, Sung, and
  Hwa]{2004DiamondSpeedofSound}
Wang,~S.-F.; Hsu,~Y.-F.; Pu,~J.-C.; Sung,~J.~C.; Hwa,~L. \emph{Materials
  Chemistry and Physics} \textbf{2004}, \emph{85}, 432--437\relax
\mciteBstWouldAddEndPuncttrue
\mciteSetBstMidEndSepPunct{\mcitedefaultmidpunct}
{\mcitedefaultendpunct}{\mcitedefaultseppunct}\relax
\EndOfBibitem
\bibitem[Ulman(1996)]{1996RevUlmanSAM}
Ulman,~A. \emph{Chemical Reviews} \textbf{1996}, \emph{96}, 1533--1554\relax
\mciteBstWouldAddEndPuncttrue
\mciteSetBstMidEndSepPunct{\mcitedefaultmidpunct}
{\mcitedefaultendpunct}{\mcitedefaultseppunct}\relax
\EndOfBibitem
\bibitem[Rego and Kirczenow(1998)Rego, and Kirczenow]{1998QuantThermCond}
Rego,~L. G.~C.; Kirczenow,~G. \emph{Phys. Rev. Lett.} \textbf{1998}, \emph{81},
  232--235\relax
\mciteBstWouldAddEndPuncttrue
\mciteSetBstMidEndSepPunct{\mcitedefaultmidpunct}
{\mcitedefaultendpunct}{\mcitedefaultseppunct}\relax
\EndOfBibitem
\bibitem[Angelescu \latin{et~al.}(1998)Angelescu, Cross, and
  Roukes]{1998HeatTranspRev}
Angelescu,~D.; Cross,~M.; Roukes,~M. \emph{Superlattices and Microstructures}
  \textbf{1998}, \emph{23}, 673--689\relax
\mciteBstWouldAddEndPuncttrue
\mciteSetBstMidEndSepPunct{\mcitedefaultmidpunct}
{\mcitedefaultendpunct}{\mcitedefaultseppunct}\relax
\EndOfBibitem
\bibitem[Schwab \latin{et~al.}(2000)Schwab, Henriksen, Worlock, and
  Roukes]{2000NatQuantThermCond}
Schwab,~K.; Henriksen,~E.~A.; Worlock,~J.~M.; Roukes,~M.~L. \emph{Nature}
  \textbf{2000}, \emph{404}, 974--977\relax
\mciteBstWouldAddEndPuncttrue
\mciteSetBstMidEndSepPunct{\mcitedefaultmidpunct}
{\mcitedefaultendpunct}{\mcitedefaultseppunct}\relax
\EndOfBibitem
\bibitem[Witt(2004)]{2004TechMolRev}
Witt,~D. \emph{Current Organic Chemistry} \textbf{2004}, \emph{8},
  1763--1797\relax
\mciteBstWouldAddEndPuncttrue
\mciteSetBstMidEndSepPunct{\mcitedefaultmidpunct}
{\mcitedefaultendpunct}{\mcitedefaultseppunct}\relax
\EndOfBibitem
\bibitem[Wang \latin{et~al.}(2007)Wang, Carter, Lagutchev, Koh, Seong, Cahill,
  and Dlott]{DlottScience07}
Wang,~Z.; Carter,~J.~A.; Lagutchev,~A.; Koh,~Y.~K.; Seong,~N.-H.;
  Cahill,~D.~G.; Dlott,~D.~D. \emph{Science} \textbf{2007}, \emph{317},
  787--790\relax
\mciteBstWouldAddEndPuncttrue
\mciteSetBstMidEndSepPunct{\mcitedefaultmidpunct}
{\mcitedefaultendpunct}{\mcitedefaultseppunct}\relax
\EndOfBibitem
\bibitem[Nitzan(2007)]{AbeScience07}
Nitzan,~A. \emph{Science} \textbf{2007}, \emph{317}, 759--760\relax
\mciteBstWouldAddEndPuncttrue
\mciteSetBstMidEndSepPunct{\mcitedefaultmidpunct}
{\mcitedefaultendpunct}{\mcitedefaultseppunct}\relax
\EndOfBibitem
\bibitem[Segal and Agarwalla(2016)Segal, and Agarwalla]{SegalReview2016}
Segal,~D.; Agarwalla,~B.~K. \emph{Annual Review of Physical Chemistry}
  \textbf{2016}, \emph{67}, 185--209, PMID: 27215814\relax
\mciteBstWouldAddEndPuncttrue
\mciteSetBstMidEndSepPunct{\mcitedefaultmidpunct}
{\mcitedefaultendpunct}{\mcitedefaultseppunct}\relax
\EndOfBibitem
\bibitem[Cui \latin{et~al.}(2017)Cui, Miao, Jiang, Meyhofer, and
  Reddy]{2017RevEnTrInMolJ}
Cui,~L.; Miao,~R.; Jiang,~C.; Meyhofer,~E.; Reddy,~P. \emph{The Journal of
  Chemical Physics} \textbf{2017}, \emph{146}, 092201\relax
\mciteBstWouldAddEndPuncttrue
\mciteSetBstMidEndSepPunct{\mcitedefaultmidpunct}
{\mcitedefaultendpunct}{\mcitedefaultseppunct}\relax
\EndOfBibitem
\bibitem[Xiong \latin{et~al.}(2018)Xiong, Xing, and Zhang]{XingThTrReview18}
Xiong,~G.; Xing,~Y.; Zhang,~L. \emph{Frontiers in Energy Research}
  \textbf{2018}, \emph{6}, 6\relax
\mciteBstWouldAddEndPuncttrue
\mciteSetBstMidEndSepPunct{\mcitedefaultmidpunct}
{\mcitedefaultendpunct}{\mcitedefaultseppunct}\relax
\EndOfBibitem
\bibitem[Rubtsov and Burin(2019)Rubtsov, and Burin]{ab19IgorReview}
Rubtsov,~I.~V.; Burin,~A.~L. \emph{The Journal of Chemical Physics}
  \textbf{2019}, \emph{150}, 020901\relax
\mciteBstWouldAddEndPuncttrue
\mciteSetBstMidEndSepPunct{\mcitedefaultmidpunct}
{\mcitedefaultendpunct}{\mcitedefaultseppunct}\relax
\EndOfBibitem
\bibitem[Qian \latin{et~al.}(2021)Qian, Zhou, and Chen]{2021PhononEngReview}
Qian,~X.; Zhou,~J.; Chen,~G. \emph{Nature Materials} \textbf{2021}, \emph{20},
  1188--1202\relax
\mciteBstWouldAddEndPuncttrue
\mciteSetBstMidEndSepPunct{\mcitedefaultmidpunct}
{\mcitedefaultendpunct}{\mcitedefaultseppunct}\relax
\EndOfBibitem
\bibitem[Anufriev \latin{et~al.}(2021)Anufriev, Wu, and Nomura]{2021ballist}
Anufriev,~R.; Wu,~Y.; Nomura,~M. \emph{Journal of Applied Physics}
  \textbf{2021}, \emph{130}, 070903\relax
\mciteBstWouldAddEndPuncttrue
\mciteSetBstMidEndSepPunct{\mcitedefaultmidpunct}
{\mcitedefaultendpunct}{\mcitedefaultseppunct}\relax
\EndOfBibitem
\bibitem[Yang \latin{et~al.}(2021)Yang, Prasher, and
  Li]{2021ReviewFromnanowtosupercond}
Yang,~L.; Prasher,~R.; Li,~D. \emph{Journal of Applied Physics} \textbf{2021},
  \emph{130}, 220901\relax
\mciteBstWouldAddEndPuncttrue
\mciteSetBstMidEndSepPunct{\mcitedefaultmidpunct}
{\mcitedefaultendpunct}{\mcitedefaultseppunct}\relax
\EndOfBibitem
\bibitem[Benenti \latin{et~al.}(2023)Benenti, Donadio, Lepri, and
  Livi]{2023RevNonFourPhTranspLiviLepri}
Benenti,~G.; Donadio,~D.; Lepri,~S.; Livi,~R. \emph{La Rivista del Nuovo
  Cimento} \textbf{2023}, \emph{46}, 105--161\relax
\mciteBstWouldAddEndPuncttrue
\mciteSetBstMidEndSepPunct{\mcitedefaultmidpunct}
{\mcitedefaultendpunct}{\mcitedefaultseppunct}\relax
\EndOfBibitem
\bibitem[Xu \latin{et~al.}(2023)Xu, Fan, and Zhou]{2023MDSimulRev}
Xu,~Y.-X.; Fan,~H.-Z.; Zhou,~Y.-G. \emph{Rare Metals} \textbf{2023}, \emph{42},
  3914--3944\relax
\mciteBstWouldAddEndPuncttrue
\mciteSetBstMidEndSepPunct{\mcitedefaultmidpunct}
{\mcitedefaultendpunct}{\mcitedefaultseppunct}\relax
\EndOfBibitem
\bibitem[Wei and Hernandez(2023)Wei, and Hernandez]{2023HeatTrModNeuralNetw}
Wei,~X.; Hernandez,~R. \emph{The Journal of Physical Chemistry Letters}
  \textbf{2023}, \emph{14}, 9834--9841\relax
\mciteBstWouldAddEndPuncttrue
\mciteSetBstMidEndSepPunct{\mcitedefaultmidpunct}
{\mcitedefaultendpunct}{\mcitedefaultseppunct}\relax
\EndOfBibitem
\bibitem[Zhang \latin{et~al.}(2024)Zhang, Zhu, Duan, Shiri, Yelishala, Shen,
  Song, Jia, Guo, Cui, and Wang]{2024EnergConvTranspRev}
Zhang,~H.; Zhu,~Y.; Duan,~P.; Shiri,~M.; Yelishala,~S.~C.; Shen,~S.; Song,~Z.;
  Jia,~C.; Guo,~X.; Cui,~L.; Wang,~K. \emph{Applied Physics Reviews}
  \textbf{2024}, \emph{11}, 041312\relax
\mciteBstWouldAddEndPuncttrue
\mciteSetBstMidEndSepPunct{\mcitedefaultmidpunct}
{\mcitedefaultendpunct}{\mcitedefaultseppunct}\relax
\EndOfBibitem
\bibitem[Liu \latin{et~al.}(2024)Liu, Wu, Zhao, Chen, Ren, Chen, and
  Zhang]{2024RevThCondControlLiu}
Liu,~C.; Wu,~C.; Zhao,~Y.; Chen,~Z.; Ren,~T.-L.; Chen,~Y.; Zhang,~G.
  \emph{Physics Reports} \textbf{2024}, \emph{1058}, 1--32\relax
\mciteBstWouldAddEndPuncttrue
\mciteSetBstMidEndSepPunct{\mcitedefaultmidpunct}
{\mcitedefaultendpunct}{\mcitedefaultseppunct}\relax
\EndOfBibitem
\bibitem[Gruebele and Wolynes(2004)Gruebele, and
  Wolynes]{2004-VibrEnReactGruebele}
Gruebele,~M.; Wolynes,~P.~G. \emph{Acc. Chem. Res.} \textbf{2004}, \emph{37},
  261--267\relax
\mciteBstWouldAddEndPuncttrue
\mciteSetBstMidEndSepPunct{\mcitedefaultmidpunct}
{\mcitedefaultendpunct}{\mcitedefaultseppunct}\relax
\EndOfBibitem
\bibitem[Botan \latin{et~al.}(2007)Botan, Backus, Pfister, Moretto, Crisma,
  Toniolo, Nguyen, Stock, and Hamm]{HammBotan07peptides}
Botan,~V.; Backus,~E. H.~G.; Pfister,~R.; Moretto,~A.; Crisma,~M.; Toniolo,~C.;
  Nguyen,~P.~H.; Stock,~G.; Hamm,~P. \emph{Proceedings of the National Academy
  of Sciences} \textbf{2007}, \emph{104}, 12749--12754\relax
\mciteBstWouldAddEndPuncttrue
\mciteSetBstMidEndSepPunct{\mcitedefaultmidpunct}
{\mcitedefaultendpunct}{\mcitedefaultseppunct}\relax
\EndOfBibitem
\bibitem[Pandey and Leitner(2016)Pandey, and
  Leitner]{PandeyLeitner2016ThermSign}
Pandey,~H.~D.; Leitner,~D.~M. \emph{The Journal of Physical Chemistry Letters}
  \textbf{2016}, \emph{7}, 5062--5067\relax
\mciteBstWouldAddEndPuncttrue
\mciteSetBstMidEndSepPunct{\mcitedefaultmidpunct}
{\mcitedefaultendpunct}{\mcitedefaultseppunct}\relax
\EndOfBibitem
\bibitem[Leitner and Yamato(2018)Leitner, and Yamato]{LeitnerReviewProtein18}
Leitner,~D.~M.; Yamato,~T. \emph{Reviews in Computational Chemistry, Volume
  31}; John Wiley and Sons, Ltd, 2018; Chapter 2, pp 63--113\relax
\mciteBstWouldAddEndPuncttrue
\mciteSetBstMidEndSepPunct{\mcitedefaultmidpunct}
{\mcitedefaultendpunct}{\mcitedefaultseppunct}\relax
\EndOfBibitem
\bibitem[Leitner(2005)]{LeitnerBook}
Leitner,~D.~M. In \emph{Geometric {S}tructures of {P}hase {S}pace in
  {M}ultidimensional {C}haos: {A}pplication to {C}hemical {R}eaction {D}ynamics
  in {C}omplex {S}stems, {PT} B}; Toda,~M., Komatsuzaki,~T., Konishi,~T.,
  Rice,~S.~A., Eds.; Adv. Chem. Phys.; John Wiley \& Sons, Inc., 2005; Vol.
  130; pp 205--256\relax
\mciteBstWouldAddEndPuncttrue
\mciteSetBstMidEndSepPunct{\mcitedefaultmidpunct}
{\mcitedefaultendpunct}{\mcitedefaultseppunct}\relax
\EndOfBibitem
\bibitem[Leitner \latin{et~al.}(2019)Leitner, Pandey, and
  Reid]{Leitner2019EnTransBiomol}
Leitner,~D.~M.; Pandey,~H.~D.; Reid,~K.~M. \emph{The Journal of Physical
  Chemistry B} \textbf{2019}, \emph{123}, 9507--9524\relax
\mciteBstWouldAddEndPuncttrue
\mciteSetBstMidEndSepPunct{\mcitedefaultmidpunct}
{\mcitedefaultendpunct}{\mcitedefaultseppunct}\relax
\EndOfBibitem
\bibitem[Elenewski \latin{et~al.}(2019)Elenewski, Velizhanin, and
  Zwolak]{Elenewski2019BiomolTr}
Elenewski,~J.~E.; Velizhanin,~K.~A.; Zwolak,~M. \emph{Nature Communications}
  \textbf{2019}, \emph{10}, 4662\relax
\mciteBstWouldAddEndPuncttrue
\mciteSetBstMidEndSepPunct{\mcitedefaultmidpunct}
{\mcitedefaultendpunct}{\mcitedefaultseppunct}\relax
\EndOfBibitem
\bibitem[Schmitz \latin{et~al.}(2019)Schmitz, Pandey, Chalyavi, Shi, Fenlon,
  Brewer, Leitner, and Tucker]{2019VibrFlReact}
Schmitz,~A.~J.; Pandey,~H.~D.; Chalyavi,~F.; Shi,~T.; Fenlon,~E.~E.;
  Brewer,~S.~H.; Leitner,~D.~M.; Tucker,~M.~J. \emph{The Journal of Physical
  Chemistry A} \textbf{2019}, \emph{123}, 10571--10581\relax
\mciteBstWouldAddEndPuncttrue
\mciteSetBstMidEndSepPunct{\mcitedefaultmidpunct}
{\mcitedefaultendpunct}{\mcitedefaultseppunct}\relax
\EndOfBibitem
\bibitem[Heyne and K{\"u}hn(2019)Heyne, and K{\"u}hn]{2019VibrReactControl}
Heyne,~K.; K{\"u}hn,~O. \emph{Journal of the American Chemical Society}
  \textbf{2019}, \emph{141}, 11730--11738\relax
\mciteBstWouldAddEndPuncttrue
\mciteSetBstMidEndSepPunct{\mcitedefaultmidpunct}
{\mcitedefaultendpunct}{\mcitedefaultseppunct}\relax
\EndOfBibitem
\bibitem[Helmer \latin{et~al.}(0)Helmer, Wolf, and
  Stock]{Stock2022EnTranspOptProt}
Helmer,~N.; Wolf,~S.; Stock,~G. \emph{The Journal of Physical Chemistry B}
  \textbf{0}, \emph{0}, null, PMID: 36261792\relax
\mciteBstWouldAddEndPuncttrue
\mciteSetBstMidEndSepPunct{\mcitedefaultmidpunct}
{\mcitedefaultendpunct}{\mcitedefaultseppunct}\relax
\EndOfBibitem
\bibitem[Leitner(2025)]{2024ThCondProtLeitner}
Leitner,~D.~M. \emph{ChemPhysChem} \textbf{2025}, \emph{26}, e202401017\relax
\mciteBstWouldAddEndPuncttrue
\mciteSetBstMidEndSepPunct{\mcitedefaultmidpunct}
{\mcitedefaultendpunct}{\mcitedefaultseppunct}\relax
\EndOfBibitem
\bibitem[Mizutani and Mizuno(2022)Mizutani, and Mizuno]{2022BarrCrossProt}
Mizutani,~Y.; Mizuno,~M. \emph{The Journal of Chemical Physics} \textbf{2022},
  \emph{157}, 240901\relax
\mciteBstWouldAddEndPuncttrue
\mciteSetBstMidEndSepPunct{\mcitedefaultmidpunct}
{\mcitedefaultendpunct}{\mcitedefaultseppunct}\relax
\EndOfBibitem
\bibitem[Rovzic \latin{et~al.}(2024)Rovzic, Teynor, Dovslic, Leitner, and
  Solomon]{2024LeitnerReact}
Rovzic,~T.; Teynor,~M.~S.; Dovslic,~N.; Leitner,~D.~M.; Solomon,~G.~C.
  \emph{Journal of Chemical Theory and Computation} \textbf{2024}, \emph{20},
  9048--9059\relax
\mciteBstWouldAddEndPuncttrue
\mciteSetBstMidEndSepPunct{\mcitedefaultmidpunct}
{\mcitedefaultendpunct}{\mcitedefaultseppunct}\relax
\EndOfBibitem
\bibitem[Ma \latin{et~al.}(2025)Ma, McCaslin, and Fournier]{2025VibrDynPept}
Ma,~Z.; McCaslin,~L.~M.; Fournier,~J.~A. \emph{Journal of the American Chemical
  Society} \textbf{2025}, \emph{147}, 9556--9565\relax
\mciteBstWouldAddEndPuncttrue
\mciteSetBstMidEndSepPunct{\mcitedefaultmidpunct}
{\mcitedefaultendpunct}{\mcitedefaultseppunct}\relax
\EndOfBibitem
\bibitem[Goldstone \latin{et~al.}(1962)Goldstone, Salam, and
  Weinberg]{1962GoldstoneSymmPart}
Goldstone,~J.; Salam,~A.; Weinberg,~S. \emph{Phys. Rev.} \textbf{1962},
  \emph{127}, 965--970\relax
\mciteBstWouldAddEndPuncttrue
\mciteSetBstMidEndSepPunct{\mcitedefaultmidpunct}
{\mcitedefaultendpunct}{\mcitedefaultseppunct}\relax
\EndOfBibitem
\bibitem[Leitner(2015)]{15LeitnerReview}
Leitner,~D.~M. \emph{Advances in Physics} \textbf{2015}, \emph{64},
  445--517\relax
\mciteBstWouldAddEndPuncttrue
\mciteSetBstMidEndSepPunct{\mcitedefaultmidpunct}
{\mcitedefaultendpunct}{\mcitedefaultseppunct}\relax
\EndOfBibitem
\bibitem[Goldstein \latin{et~al.}(2015)Goldstein, Huse, Lebowitz, and
  Tumulka]{Lebowitz15}
Goldstein,~S.; Huse,~D.~A.; Lebowitz,~J.~L.; Tumulka,~R. \emph{Phys. Rev.
  Lett.} \textbf{2015}, \emph{115}, 100402\relax
\mciteBstWouldAddEndPuncttrue
\mciteSetBstMidEndSepPunct{\mcitedefaultmidpunct}
{\mcitedefaultendpunct}{\mcitedefaultseppunct}\relax
\EndOfBibitem
\bibitem[Livi(2023)]{2021RevLiviPedagog}
Livi,~R. \emph{Physica A: Statistical Mechanics and its Applications}
  \textbf{2023}, \emph{631}, 127779, Lecture Notes of the 15th International
  Summer School of Fundamental Problems in Statistical Physics\relax
\mciteBstWouldAddEndPuncttrue
\mciteSetBstMidEndSepPunct{\mcitedefaultmidpunct}
{\mcitedefaultendpunct}{\mcitedefaultseppunct}\relax
\EndOfBibitem
\bibitem[Anderson(1958)]{Anderson58}
Anderson,~P.~W. \emph{Phys. Rev.} \textbf{1958}, \emph{109}, 1492--1505\relax
\mciteBstWouldAddEndPuncttrue
\mciteSetBstMidEndSepPunct{\mcitedefaultmidpunct}
{\mcitedefaultendpunct}{\mcitedefaultseppunct}\relax
\EndOfBibitem
\bibitem[Abrahams \latin{et~al.}(1979)Abrahams, Anderson, Licciardello, and
  Ramakrishnan]{Abrahams1979ScThLoc}
Abrahams,~E.; Anderson,~P.~W.; Licciardello,~D.~C.; Ramakrishnan,~T.~V.
  \emph{Phys. Rev. Lett.} \textbf{1979}, \emph{42}, 673--676\relax
\mciteBstWouldAddEndPuncttrue
\mciteSetBstMidEndSepPunct{\mcitedefaultmidpunct}
{\mcitedefaultendpunct}{\mcitedefaultseppunct}\relax
\EndOfBibitem
\bibitem[Rieder \latin{et~al.}(1967)Rieder, Lebowitz, and Lieb]{Lebowitz67}
Rieder,~Z.; Lebowitz,~J.~L.; Lieb,~E. \emph{Journal of Mathematical Physics}
  \textbf{1967}, \emph{8}, 1073--1078\relax
\mciteBstWouldAddEndPuncttrue
\mciteSetBstMidEndSepPunct{\mcitedefaultmidpunct}
{\mcitedefaultendpunct}{\mcitedefaultseppunct}\relax
\EndOfBibitem
\bibitem[Leitner(2001)]{20011DGlass}
Leitner,~D.~M. \emph{Phys. Rev. B} \textbf{2001}, \emph{64}, 094201\relax
\mciteBstWouldAddEndPuncttrue
\mciteSetBstMidEndSepPunct{\mcitedefaultmidpunct}
{\mcitedefaultendpunct}{\mcitedefaultseppunct}\relax
\EndOfBibitem
\bibitem[Lepri(1998)]{1998LepriAnomalThCondFPUCl}
Lepri,~S. \emph{Phys. Rev. E} \textbf{1998}, \emph{58}, 7165--7171\relax
\mciteBstWouldAddEndPuncttrue
\mciteSetBstMidEndSepPunct{\mcitedefaultmidpunct}
{\mcitedefaultendpunct}{\mcitedefaultseppunct}\relax
\EndOfBibitem
\bibitem[Narayan and Ramaswamy(2002)Narayan, and Ramaswamy]{Turbulence02}
Narayan,~O.; Ramaswamy,~S. \emph{Phys. Rev. Lett.} \textbf{2002}, \emph{89},
  200601\relax
\mciteBstWouldAddEndPuncttrue
\mciteSetBstMidEndSepPunct{\mcitedefaultmidpunct}
{\mcitedefaultendpunct}{\mcitedefaultseppunct}\relax
\EndOfBibitem
\bibitem[Burin(2025)]{ab25SuperDiffPh}
Burin,~A.~L. \emph{The Journal of Chemical Physics} \textbf{2025}, \emph{162},
  165102\relax
\mciteBstWouldAddEndPuncttrue
\mciteSetBstMidEndSepPunct{\mcitedefaultmidpunct}
{\mcitedefaultendpunct}{\mcitedefaultseppunct}\relax
\EndOfBibitem
\bibitem[Lee \latin{et~al.}(2017)Lee, Wu, Lou, Lee, and
  Chang]{2017cntHighKappa_mmlong}
Lee,~V.; Wu,~C.-H.; Lou,~Z.-X.; Lee,~W.-L.; Chang,~C.-W. \emph{Phys. Rev.
  Lett.} \textbf{2017}, \emph{118}, 135901\relax
\mciteBstWouldAddEndPuncttrue
\mciteSetBstMidEndSepPunct{\mcitedefaultmidpunct}
{\mcitedefaultendpunct}{\mcitedefaultseppunct}\relax
\EndOfBibitem
\bibitem[Maire \latin{et~al.}(2017)Maire, Anufriev, and
  Nomura]{2017BallTr4KSiNanow033}
Maire,~J.; Anufriev,~R.; Nomura,~M. \emph{Scientific Reports} \textbf{2017},
  \emph{7}, 41794\relax
\mciteBstWouldAddEndPuncttrue
\mciteSetBstMidEndSepPunct{\mcitedefaultmidpunct}
{\mcitedefaultendpunct}{\mcitedefaultseppunct}\relax
\EndOfBibitem
\bibitem[Meier \latin{et~al.}(2014)Meier, Menges, Nirmalraj, H\"olscher, Riel,
  and Gotsmann]{MeierMolChains14}
Meier,~T.; Menges,~F.; Nirmalraj,~P.; H\"olscher,~H.; Riel,~H.; Gotsmann,~B.
  \emph{Phys. Rev. Lett.} \textbf{2014}, \emph{113}, 060801\relax
\mciteBstWouldAddEndPuncttrue
\mciteSetBstMidEndSepPunct{\mcitedefaultmidpunct}
{\mcitedefaultendpunct}{\mcitedefaultseppunct}\relax
\EndOfBibitem
\bibitem[Liu \latin{et~al.}(2024)Liu, Jhalaria, Ruzicka, Benicewicz, Kumar,
  Fytas, and Xu]{2024prlnanopartonpolymsuperdiff}
Liu,~B.; Jhalaria,~M.; Ruzicka,~E.; Benicewicz,~B.~C.; Kumar,~S.~K.; Fytas,~G.;
  Xu,~X. \emph{Phys. Rev. Lett.} \textbf{2024}, \emph{133}, 248101\relax
\mciteBstWouldAddEndPuncttrue
\mciteSetBstMidEndSepPunct{\mcitedefaultmidpunct}
{\mcitedefaultendpunct}{\mcitedefaultseppunct}\relax
\EndOfBibitem
\bibitem[Flory(1969)]{1969florystatistical}
Flory,~P. \emph{Statistical Mechanics of Chain Molecules}; Interscience
  Publishers, 1969\relax
\mciteBstWouldAddEndPuncttrue
\mciteSetBstMidEndSepPunct{\mcitedefaultmidpunct}
{\mcitedefaultendpunct}{\mcitedefaultseppunct}\relax
\EndOfBibitem
\bibitem[Rubinstein and Colby(2003)Rubinstein, and
  Colby]{2003rubinsteinpolymer}
Rubinstein,~M.; Colby,~R. \emph{Polymer Physics}; OUP Oxford, 2003\relax
\mciteBstWouldAddEndPuncttrue
\mciteSetBstMidEndSepPunct{\mcitedefaultmidpunct}
{\mcitedefaultendpunct}{\mcitedefaultseppunct}\relax
\EndOfBibitem
\bibitem[Duan \latin{et~al.}(2019)Duan, Li, Liu, Chen, and
  Li]{2019ThermCondKinksXuhui}
Duan,~X.; Li,~Z.; Liu,~J.; Chen,~G.; Li,~X. \emph{Journal of Applied Physics}
  \textbf{2019}, \emph{125}, 164303\relax
\mciteBstWouldAddEndPuncttrue
\mciteSetBstMidEndSepPunct{\mcitedefaultmidpunct}
{\mcitedefaultendpunct}{\mcitedefaultseppunct}\relax
\EndOfBibitem
\bibitem[Dinpajooh and Nitzan(2020)Dinpajooh, and
  Nitzan]{2020NitzanHeatCondMDRegimes}
Dinpajooh,~M.; Nitzan,~A. \emph{The Journal of Chemical Physics} \textbf{2020},
  \emph{153}, 164903\relax
\mciteBstWouldAddEndPuncttrue
\mciteSetBstMidEndSepPunct{\mcitedefaultmidpunct}
{\mcitedefaultendpunct}{\mcitedefaultseppunct}\relax
\EndOfBibitem
\bibitem[Frisch \latin{et~al.}(2009)Frisch, Trucks, Schlegel, Scuseria, Robb,
  Cheeseman, Scalmani, Barone, Mennucci, Petersson, Nakatsuji, Caricato, Li,
  Hratchian, Izmaylov, Bloino, Zheng, Sonnenberg, Hada, Ehara, Toyota, Fukuda,
  Hasegawa, Ishida, Nakajima, Honda, Kitao, Nakai, Vreven, Montgomery, Peralta,
  Ogliaro, Bearpark, Heyd, Brothers, Kudin, Staroverov, Kobayashi, Normand,
  Raghavachari, Rendell, Burant, Iyengar, Tomasi, Cossi, Rega, Millam, Klene,
  Knox, Cross, Bakken, Adamo, Jaramillo, Gomperts, Stratmann, Yazyev, Austin,
  Cammi, Pomelli, Ochterski, Martin, Morokuma, Zakrzewski, Voth, Salvador,
  Dannenberg, Dapprich, Daniels, Farkas, Foresman, Ortiz, Cioslowski, and
  Fox]{Gaussian}
Frisch,~M.~J. \latin{et~al.}  Gaussian~09 {R}evision {A}.1. 2009; Gaussian Inc.
  Wallingford CT 2009\relax
\mciteBstWouldAddEndPuncttrue
\mciteSetBstMidEndSepPunct{\mcitedefaultmidpunct}
{\mcitedefaultendpunct}{\mcitedefaultseppunct}\relax
\EndOfBibitem
\bibitem[Landau \latin{et~al.}(1986)Landau, Lifshitz, Kosevich, Sykes,
  Pitaevskii, and Reid]{landau1986Elasticitytheory}
Landau,~L.; Lifshitz,~E.; Kosevich,~A.; Sykes,~J.; Pitaevskii,~L.; Reid,~W.
  \emph{Theory of Elasticity: Volume 7}; Course of theoretical physics;
  Elsevier Science, 1986\relax
\mciteBstWouldAddEndPuncttrue
\mciteSetBstMidEndSepPunct{\mcitedefaultmidpunct}
{\mcitedefaultendpunct}{\mcitedefaultseppunct}\relax
\EndOfBibitem
\bibitem[Burin \latin{et~al.}(2023)Burin, Parshin, and
  Rubtsov]{ab2023Cherenkov}
Burin,~A.~L.; Parshin,~I.~V.; Rubtsov,~I.~V. \emph{The Journal of Chemical
  Physics} \textbf{2023}, \emph{159}, 054903\relax
\mciteBstWouldAddEndPuncttrue
\mciteSetBstMidEndSepPunct{\mcitedefaultmidpunct}
{\mcitedefaultendpunct}{\mcitedefaultseppunct}\relax
\EndOfBibitem
\bibitem[Segal \latin{et~al.}(2003)Segal, Nitzan, and H\"anggi]{SegalNitzan03}
Segal,~D.; Nitzan,~A.; H\"anggi,~P. \emph{The Journal of Chemical Physics}
  \textbf{2003}, \emph{119}, 6840--6855\relax
\mciteBstWouldAddEndPuncttrue
\mciteSetBstMidEndSepPunct{\mcitedefaultmidpunct}
{\mcitedefaultendpunct}{\mcitedefaultseppunct}\relax
\EndOfBibitem
\bibitem[Landauer(1957)]{Landauer57Classic}
Landauer,~R. \emph{IBM Journal of Research and Development} \textbf{1957},
  \emph{1}, 223--231\relax
\mciteBstWouldAddEndPuncttrue
\mciteSetBstMidEndSepPunct{\mcitedefaultmidpunct}
{\mcitedefaultendpunct}{\mcitedefaultseppunct}\relax
\EndOfBibitem
\bibitem[Klitsner \latin{et~al.}(1988)Klitsner, VanCleve, Fischer, and
  Pohl]{1988EquatForThermCond}
Klitsner,~T.; VanCleve,~J.~E.; Fischer,~H.~E.; Pohl,~R.~O. \emph{Phys. Rev. B}
  \textbf{1988}, \emph{38}, 7576--7594\relax
\mciteBstWouldAddEndPuncttrue
\mciteSetBstMidEndSepPunct{\mcitedefaultmidpunct}
{\mcitedefaultendpunct}{\mcitedefaultseppunct}\relax
\EndOfBibitem
\bibitem[Dwivedi and Chua(2016)Dwivedi, and Chua]{2016GeneralizedTransfMatr1D}
Dwivedi,~V.; Chua,~V. \emph{Phys. Rev. B} \textbf{2016}, \emph{93},
  134304\relax
\mciteBstWouldAddEndPuncttrue
\mciteSetBstMidEndSepPunct{\mcitedefaultmidpunct}
{\mcitedefaultendpunct}{\mcitedefaultseppunct}\relax
\EndOfBibitem
\bibitem[Caroli \latin{et~al.}(1971)Caroli, Combescot, Lederer, Nozieres, and
  Saint-James]{1971TrThrGFCaroli}
Caroli,~C.; Combescot,~R.; Lederer,~D.; Nozieres,~P.; Saint-James,~D.
  \emph{Journal of Physics C: Solid State Physics} \textbf{1971}, \emph{4},
  2598\relax
\mciteBstWouldAddEndPuncttrue
\mciteSetBstMidEndSepPunct{\mcitedefaultmidpunct}
{\mcitedefaultendpunct}{\mcitedefaultseppunct}\relax
\EndOfBibitem
\bibitem[Mujica \latin{et~al.}(1994)Mujica, Kemp, and Ratner]{1994MWCoductMR}
Mujica,~V.; Kemp,~M.; Ratner,~M.~A. \emph{The Journal of Chemical Physics}
  \textbf{1994}, \emph{101}, 6849--6855\relax
\mciteBstWouldAddEndPuncttrue
\mciteSetBstMidEndSepPunct{\mcitedefaultmidpunct}
{\mcitedefaultendpunct}{\mcitedefaultseppunct}\relax
\EndOfBibitem
\bibitem[Galperin \latin{et~al.}(2004)Galperin, Ratner, and
  Nitzan]{2004TransmGFRatner}
Galperin,~M.; Ratner,~M.~A.; Nitzan,~A. \emph{The Journal of Chemical Physics}
  \textbf{2004}, \emph{121}, 11965--11979\relax
\mciteBstWouldAddEndPuncttrue
\mciteSetBstMidEndSepPunct{\mcitedefaultmidpunct}
{\mcitedefaultendpunct}{\mcitedefaultseppunct}\relax
\EndOfBibitem
\bibitem[Wang \latin{et~al.}(2008)Wang, Wang, and
  L{\"u}]{2008QuantThermTranspRev}
Wang,~J.-S.; Wang,~J.; L{\"u},~J.~T. \emph{The European Physical Journal B}
  \textbf{2008}, \emph{62}, 381--404\relax
\mciteBstWouldAddEndPuncttrue
\mciteSetBstMidEndSepPunct{\mcitedefaultmidpunct}
{\mcitedefaultendpunct}{\mcitedefaultseppunct}\relax
\EndOfBibitem
\bibitem[MATLAB(2019)]{MATLAB:2019}
MATLAB \emph{version 9.6.0 (R2019a)}; The MathWorks Inc.: Natick,
  Massachusetts, 2019\relax
\mciteBstWouldAddEndPuncttrue
\mciteSetBstMidEndSepPunct{\mcitedefaultmidpunct}
{\mcitedefaultendpunct}{\mcitedefaultseppunct}\relax
\EndOfBibitem
\bibitem[Dorokhov(1984)]{1984DOROKHOV381}
Dorokhov,~O. \emph{Solid State Communications} \textbf{1984}, \emph{51},
  381--384\relax
\mciteBstWouldAddEndPuncttrue
\mciteSetBstMidEndSepPunct{\mcitedefaultmidpunct}
{\mcitedefaultendpunct}{\mcitedefaultseppunct}\relax
\EndOfBibitem
\bibitem[Islam \latin{et~al.}(2023)Islam, Nawagamuwage, Parshin, Richard,
  Burin, and Rubtsov]{ab2023Membr}
Islam,~M.~M.; Nawagamuwage,~S.~U.; Parshin,~I.~V.; Richard,~M.~C.;
  Burin,~A.~L.; Rubtsov,~I.~V. \emph{Journal of the American Chemical Society}
  \textbf{2023}, \emph{145}, 26363--26373\relax
\mciteBstWouldAddEndPuncttrue
\mciteSetBstMidEndSepPunct{\mcitedefaultmidpunct}
{\mcitedefaultendpunct}{\mcitedefaultseppunct}\relax
\EndOfBibitem
\end{mcitethebibliography}

\end{document}